\documentclass[12pt]{article}

\usepackage[utf8]{inputenc}
\usepackage[english]{babel}
\usepackage[a4paper, left=2.5cm, right=2.5cm, top=2.5cm, bottom=2.5cm]{geometry}
\usepackage{setspace}
\usepackage{graphicx}
\usepackage{caption}
\usepackage{float}
\usepackage{fancyhdr}
\usepackage{lineno} 
\usepackage{amsmath, amssymb}
\usepackage{booktabs}
\usepackage{hyperref}
\hypersetup{colorlinks=true, linkcolor=black, citecolor=black, urlcolor=blue}
\usepackage{titlesec}
\usepackage{array} 
\usepackage{array}       
\usepackage{tabularx}    
\usepackage{booktabs}    
\usepackage{makecell}
\usepackage[utf8]{inputenc}
\usepackage{url}
\title{\textbf{Genotype–Phenotype Integration through Machine Learning and\\
Personalised Gene Regulatory Networks for Cancer Metastasis Prediction}}

\date{} 
\title{\textbf{Genotype–Phenotype Integration through Machine Learning and\\
Personalised Gene Regulatory Networks for Cancer Metastasis Prediction}}

\author{
Jiwei Fu$^{1,*}$ \quad Chunyu Yang$^{1}$ \\[6pt]
\small $^{1}$Nuffield Department of Medicine, University of Oxford \\[3pt]
\small $^{*}$Corresponding author
}

\date{}

\begin{document}

\maketitle

\section*{Abstract}
\addcontentsline{toc}{section}{Abstract}

\textbf{Background and Aims} \\
Metastasis is the leading cause of cancer-related mortality. Most predictive models focus on single cancer types and overlook patient-specific regulatory networks. This study had two objectives. The first was to evaluate expression-based machine learning (ML) models for pancancer metastasis prediction and assess their feasibility in resource-limited healthcare settings. The second was to construct personalised gene regulatory networks and apply a graph neural network (GNN) to capture patient-level regulatory patterns. \\[0.5em]
\textbf{Methodology and Key Findings} \\
Gene expression data were obtained from the Cancer Cell Line Encyclopedia. A transcription factor--target prior from DoRothEA was restricted to nine metastasis-associated transcription factors. Differential genes were selected using the Kruskal-Wallis test. ElasticNet, Random Forest, and XGBoost were trained on top-ranked genes. The personalised networks were generated with PANDA and LIONESS, and analysed using a GATv2 model with topological and expression features. XGBoost with 1,000 genes achieved the highest ML performance (AUROC 0.7051, MCC 0.2914). The best GNN model used 100 genes (AUROC 0.6423, MCC 0.2254), showing similar sensitivity for metastatic cases but lower overall accuracy. PCA and graph-level statistics indicated limited intrinsic separability, and both approaches were robust to data partitioning. \\[0.5em]
\textbf{Conclusion and Significance} \\
Expression-based ML shows potential as a cost-efficient approach for preliminary screening in resource-limited contexts, although current performance remains moderate. Personalised GNN modelling enables patient-specific network analysis, though its effectiveness is constrained by weak topology signals in this dataset. Combining ML-based benchmarking with patient-level network analysis can guide resource allocation in different healthcare settings and support precision cancer care.

\section{Introduction}
Cancer causes nearly 10 million deaths annually, with metastasis the leading cause \cite{who_cancer}. Low- and middle-income countries (LMICs) bear about 70\% of this burden, and the incidence is projected to increase substantially in the coming decades \cite{ascopubs_lmic}. Limited resources in these settings hinder the adoption of early diagnosis and personalised treatment. Reliable early detection and precision strategies are therefore essential to improve survival. The growing accessibility of high-throughput sequencing and artificial intelligence has accelerated research on cancer metastases, with numerous machine learning and deep learning models now being developed to enable early detection and personalised treatment\cite{frontiers_ai_cancer}.

Machine learning (ML) approaches have demonstrated a strong capability in predicting cancer metastasis status from differential gene expression profiles, and models such as XGBoost achieve high accuracy in breast cancer metastasis classification \cite{li2022xgboost,jung2023breast}.
More recently, graph neural networks (GNNs) have been used with biological information to improve metastasis diagnosis and prediction \cite{kipf2017semi,hamilton2017inductive}.  For example, a study integrated multi-omics data into a GNN framework to identify cancer gene modules linked to metastatic progression \cite{emogi2021nmi}.In another similar study, a GNN was applied to biological networks, and interpretability techniques were used to identify functional gene modules that can be associated with metastasis \cite{schulte2021glrp}. In addition, a study used a GNN that took gene expression data and protein–protein interaction networks as input to predict metastasis and survival risk \cite{gnnsubnet2022bioinformatics}. Besides, an attention-based GNN was used to find cancer gene modules linked to metastatic progression. \cite{cgmmega2023}.  In addition, a multitask GNN model improved metastasis prediction by learning cancer driver gene classification at the same time, so it could use useful information from both tasks to make better predictions \cite{mtgcn2022bib}.In breast cancer, GNN models using ultrasound and histopathology data improved the precision of lymph node metastasis detection \cite{breastgnn2025}. In non-small cell lung cancer, a GNN that integrates PET / CT characteristics with protein-protein interaction networks improved the prediction of metastasis risk \cite{nsclcgnn2024}. For colorectal cancer, spatio-temporal GNNs that integrate multimodal data achieved early prediction of liver metastases \cite{crlmgnn2025}.

The research gap mainly lies in two areas. First, most existing ML models predict metastasis status only for a single cancer type. \cite{li2022xgboost,jung2023breast}. Few studies have systematically assessed these methods for the prediction of pancancer metastases and evaluated their applicability in low- and middle-income countries (LMICs).
Second, in GNN-based biomedical studies, networks are typically static and shared by all samples, which ignores the patient-specific network structure. PANDA and LIONESS \cite{panda, lioness}are two algorithms designed to address this problem. This study addresses both gaps by using CCLE \cite{ccle} for large-scale pancancer evaluation of ML models and assessing their potential applicability in low- and middle-income countries (LMICs). Furthermore, the PANDA and LIONESS algorithms are applied to construct personalised networks for each sample in GNN-based metastasis classification.

Building on these gaps, this study has three  goals:  
\begin{enumerate}
    \item Use the CCLE gene expression data to test how well common machine learning models predict pancancer metastasis, and assess their potential applicability in low- and middle-income countries (LMICs).
    \item Build a personalised network for each sample using PANDA and LIONESS, apply these networks in a GNN model for metastasis prediction, and assess its performance.
    \item Use the machine learning models from Goal~1 as benchmarks to compare with the GNN models.
\end{enumerate}

\section{ Methods}

\subsection{Data}
\subsubsection{Data Sources}

This study used gene expression data from the Cancer Cell Line Encyclopaedia (CCLE) \cite{ccle}, which measures how active each gene is in different cancer cell lines. It also used a transcription factor–target prior (TF--target)from the DoRothEA database \cite{garcia2019dorothea}, which is a reference list showing which transcription factors (proteins that regulate the activity of specific genes) control which genes. For this study, the list was filtered to nine transcription factors linked to metastasis (TP53, MYC, STAT3, HIF1A, NFKB1, SOX2, TWIST1, SNAI1, and ZEB1\cite{yang2004,batlle2000,spaderna2008,yu2009,semenza2002,boumahdi2020,cooper_ras,muller2009,snail_nfkb}). Expression data annotated with each sample's metastatic status was used to train machine learning models for performance evaluation and TF--target data were combined with gene expression data to generate a gene regulatory network (GRN) for each sample.

\subsubsection{Data Processing}

Data processing involved three stages. Metastatic status was added to the sample identifiers in the expression matrix. The distribution of metastatic status was examined. A balanced data set was then created with an equal number of primary and metastatic samples. Using the expression matrix, the top 100, 200, 500, and 1000 gene features were selected according to the Kruskal statistical test \cite{kruskal1952}.

\begin{figure}[H]
    \centering
    \includegraphics[width=0.75\textwidth]{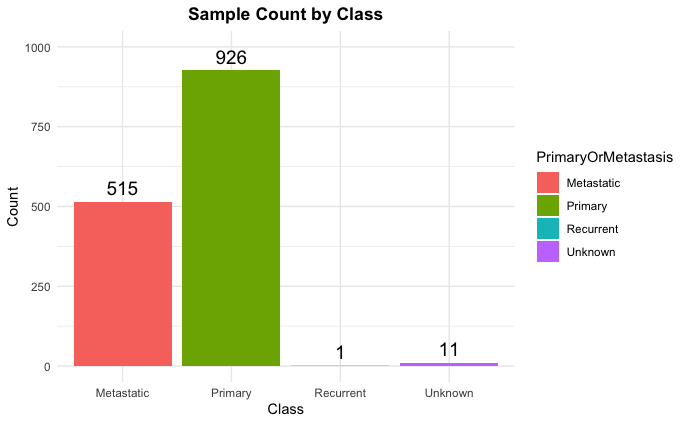}\\[1em]
    \includegraphics[width=0.75\textwidth]{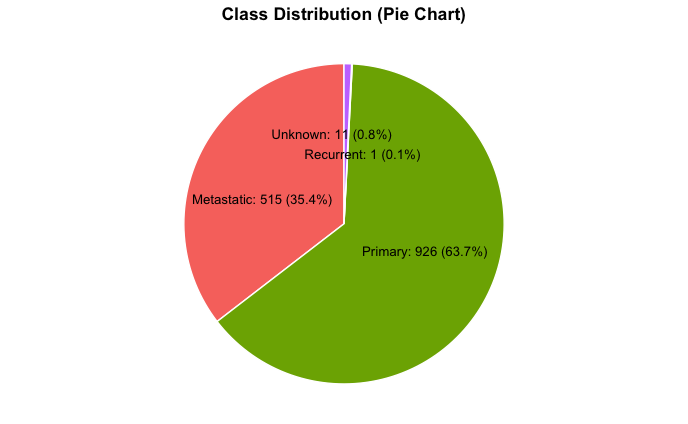}
    \caption{Sample count by class and class distribution. 
    The bar plot (top) and pie chart (bottom) show that the dataset contains 926 primary samples (63.7\%), 515 metastatic samples (35.4\%), 1 recurrent sample (0.1\%), and 11 samples with unknown classification (0.8\%).}
    \label{fig:class_distribution}
\end{figure}

The distribution of metastatic status before balancing is shown in Figure~\ref{fig:class_distribution}, which illustrates the predominance of primary samples, with very few recurrent or unknown cases.

This imbalance can bias the outcome toward the majority class due to the overfitting problem. Creating a balanced data set mitigates this risk \cite{chawla2002smote,he2009learning}. Multiple top-$N$ gene subsets (100, 200, 500, 1000) were evaluated to identify the subset that produces the best performance, as commonly practised in previous studies.

\subsubsection{Exploratory analysis }

The exploratory analysis included a volcano plot and a heatmap. The X-axis shows the average expression difference between metastatic and primary samples, and the Y-axis shows the statistical significance from the Kruskal test. This plot was used to see if many genes are consistently higher or lower in metastatic samples compared to primary samples. A heat map of the 50 most statistically important genes was visualised to assess whether these genes could separate the two phenotypes.

\subsection{ Methodology}
\subsubsection{Workflow Overview} 
\begin{figure}[H]
    \centering
    \includegraphics[width=0.75\textwidth]{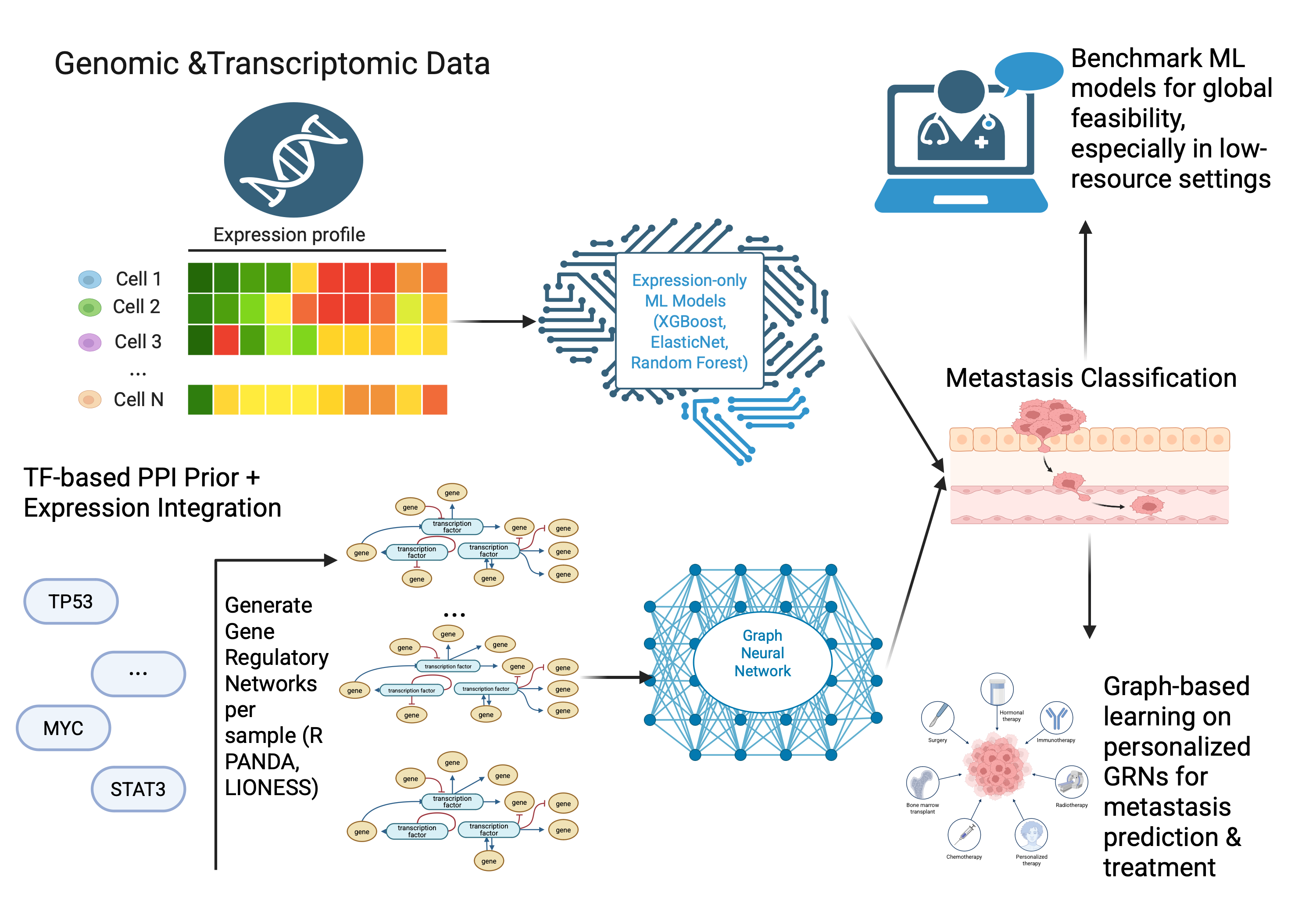}
    \caption{Workflow of the proposed analysis framework. Gene expression profiles were analyzed using two complementary approaches: (1) expression-only machine learning models (XGBoost, ElasticNet, Random Forest) for metastasis classification, serving both as benchmarks for GNN performance and for assessing feasibility in low-resource settings;
     (2) Graph neural networks are trained on personalized gene regulatory networks. These networks are generated by integrating TF-target data with expression data using the PANDA and LIONESS algorithms.}
    \label{fig:workflow}
\end{figure}

Figure~\ref{fig:workflow} illustrates the workflow of the proposed analysis framework.

\subsubsection{Traditional Machine Learning Models}
This study employs three machine learning models that are used for the classification of metastases and to assess feasibility in low-resource settings. These models were selected due to their strong performance in gene expression-based classification tasks\cite{piccolo2022classify}. Table~\ref{tab:ml_models} summarises their core mechanisms and the rationale for their suitability within the proposed analysis pipeline.

\begin{table}[h]
\centering

\begin{tabular}{p{3cm} p{4cm} p{9cm}}
\toprule
\textbf{Model} & \textbf{Algorithm Type} & \textbf{Core Mechanism} \\
\midrule
ElasticNet  & Linear model with regularization & A linear model that selects important genes and reduces the effect of less relevant ones. $L_1$ regularization encourages sparsity by setting some coefficients to zero. $L_2$ regularization stabilizes the model by shrinking large coefficients.\cite{zou2005regularization} \\
Random Forest  & Ensemble method (bagging) & An ensemble method that builds many decision trees from resampled gene expression data and combines their predictions. It captures complex gene–gene relationships, reduces overfitting, and is robust to noise.\cite{breiman2001random}\\

XGBoost & Ensemble method (gradient boosting) & An ensemble method that builds decision trees sequentially, where each tree corrects the errors of the previous ones using gradient boosting. It handles sparse gene expression data efficiently and remains computationally efficient.  \cite{chen2016xgboost}\\
\bottomrule
\end{tabular}
\caption{Summary of traditional machine learning models used for metastasis classification.}
\label{tab:ml_models}
\end{table}

\subsubsection{GRN Construction}

In this study, the construction of the Gene regulatory networks involved three main steps. First, gene expression data were integrated with prior transcription factor TF–target interactions to infer a consensus regulatory network using the PANDA algorithm. The TF–target prior was restricted to nine transcription factors with established roles in cancer metastasis: TWIST1 \cite{yang2004}, SNAI1 \cite{batlle2000}, ZEB1 \cite{spaderna2008}, STAT3 \cite{yu2009}, HIF1A \cite{semenza2002}, SOX2 \cite{boumahdi2020}, MYC \cite{cooper_ras}, TP53 \cite{muller2009}, and NFKB1 \cite{snail_nfkb}, which regulate key processes such as motility, invasion, angiogenesis, stemness, immune evasion, and tumour microenvironment remodelling. Second, sample-specific GRNs were generated by applying the LIONESS framework, which estimates the individual contribution of each sample to the overall network. Finally, each LIONESS-derived network was converted into a graph object in PyTorch Geometric (PyG) format for input into the graph neural network (GNN) model.

The following describes the PANDA and LIONESS algorithms for constructing consensus and sample-specific gene regulatory networks.

\paragraph{PANDA network}

PANDA (Passing Attributes Between Networks for Data Assimilation) integrates a transcription factor-target (TF-target) and gene expression data to produce a consensus gene regulatory network (GRN)\cite{panda} in three steps:

\begin{table}[H]
\centering
\caption{Three-step iterative procedure of the PANDA algorithm.}
\begin{tabular}{p{3.5cm} p{6.5cm} p{5cm}}
\toprule
\textbf{Step} & \textbf{Description} & \textbf{Formula} \\
\midrule
\textbf{1: Responsibility} & Measures how well a TF–gene link is supported by the TF’s cooperativity partners. &
$R_{ij} = z\!\Big( \sum_{k} P_{ik}\, W_{kj} \Big)$ \\

\textbf{2: Availability} & Measures whether a target gene is co-expressed with other genes regulated by the same TF. In other words, if a transcription factor regulates multiple genes, this value checks whether the target gene shows similar expression patterns to those other genes across samples. &
$A_{ij} = z\!\Big( \sum_{k} W_{ik}\, C_{kj} \Big)$ \\

\textbf{3: Edge update} & Combines $R$ and $A$ to refine TF–gene link strength; repeat until $W$ converges. &
$W^{(t+1)}_{ij} = (1-\alpha) W^{(t)}_{ij} + \alpha \cdot \frac{R_{ij} + A_{ij}}{2}$ \\
\midrule
\multicolumn{3}{p{15cm}}{\textbf{Notation:} 
$P$: TF–TF cooperativity network (identity matrix in this study); 
$W$: TF–target weight matrix; 
$C$: gene–gene co-expression network from expression data; 
$z(\cdot)$: z-score normalization; 
$\alpha$: update step size; 
$R$: responsibility score; 
$A$: availability score.} \\
\bottomrule
\end{tabular}
\label{tab:panda_steps}
\end{table}

The advantage of PANDA is that it produces a GRN that reflects both known biology and coexpression patterns throughout the sample.\cite{panda} The network was then used as input for LIONESS to estimate sample-specific GRNs. The regulatory subgraph TP53 with the top 50 genes of PANDA is shown in the results for illustration.

\begin{figure}[H]
    \centering
    \includegraphics[width=0.85\textwidth]{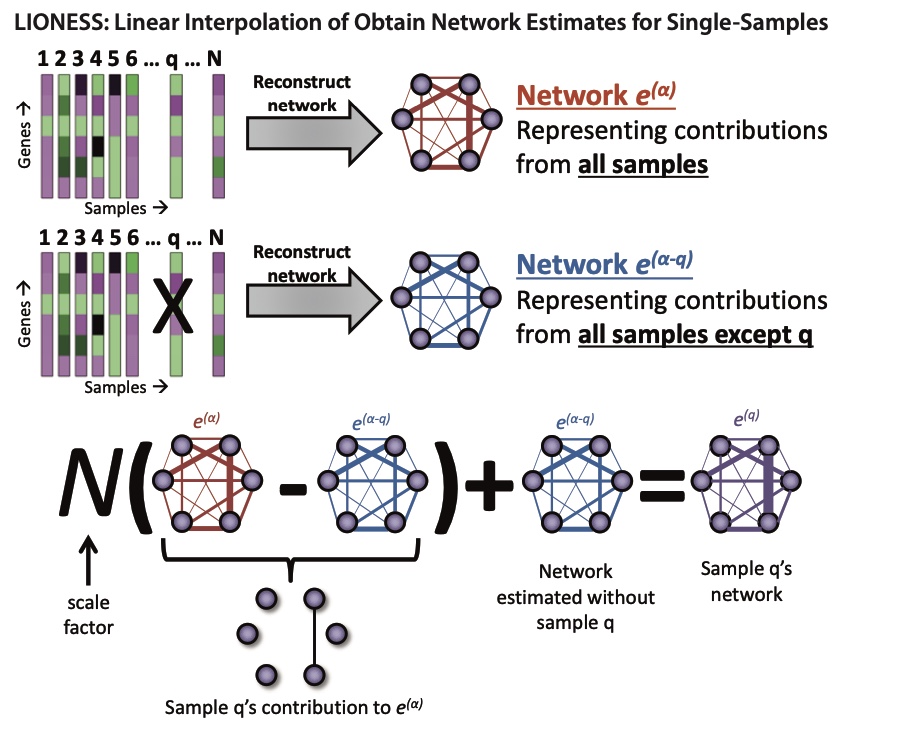}
    \caption{\textbf{Overview of the LIONESS framework.} 
     Figure adapted from Kuijjer \textit{et al.} (2019)\cite{lioness}.}
    \label{fig:lioness}
\end{figure}

\paragraph{Lioness network}
LIONESS was applied to transform the PANDA network into sample-specific gene regulatory networks. The method reconstructs a network using all samples ($e^{(\alpha)}$) and again with one sample $q$ removed ($e^{(\alpha-q)}$). Comparing the network built with all samples to the network built without sample $q$, and then scaling this difference by the total number of samples $N$, gives the contribution of sample $q$ and results in its individual GRN (Figure~\ref{fig:lioness}). The advantage of using LIONESS is its ability to infer sample-specific gene regulatory networks while preserving the global structure, thereby providing a more accurate representation of individual-specific regulatory patterns.

The LIONESS equation for estimating the sample-specific network is:
\begin{equation}
    e^{(q)} = N \left( e^{(\alpha)} - e^{(\alpha-q)} \right) + e^{(\alpha-q)},
\end{equation}
where:
\begin{itemize}
    \item $e^{(\alpha)}$: network built from all samples,
    \item $e^{(\alpha-q)}$: network built from all samples except sample $q$,
    \item $N$: total number of samples,
    \item $e^{(q)}$: estimated network for sample $q$.
\end{itemize}


\subsubsection{Graph Neural Networks (GNNs)}

The sample-specific gene regulatory networks (GRNs) were used for graph neural network (GNN) analysis. The use of GNNs allows the model to capture complex topological patterns within regulatory networks \cite{kipf2017semi,hamilton2017inductive} that can be associated with metastatic status.

Table~\ref{tab:gnn_graph_structure} summarises the graph used for GNN analysis. 
The features of the nodes include degree and centrality of kinship, which encode local connectivity and potential control over the information flow (i.e., the ability of a node to act as a bridge by connecting otherwise distant parts of the network) \cite{freeman1977,newman2010}. 
Gene expression values are z-score normalised across samples to place features on a comparable scale and stabilise optimisation \cite{bishop2006prml}. 
A node role indicator (TF / Target / TF+Target) provides biological context to guide message passing. 
Additional features were not included, as the analysis aims to test whether regulatory topology alone carries a signal for metastasis.

\renewcommand{\arraystretch}{1.3}
\begin{table}[ht]
\centering
\caption{Graph structure used for GNN analysis.}
\begin{tabular}{p{3cm} p{9cm}}
\toprule
\textbf{Component} & \textbf{Description} \\
\midrule
\textbf{Nodes} & Each node contains four features: degree, betweenness centrality, normalised gene expression value, and node role (TF / Target / TF+Target). \\
\textbf{Edges} &  Edge weights reflect regulatory strength. \\
\textbf{Graph type} & Directed, weighted graph. \\

\bottomrule
\end{tabular}

\noindent\rule{\linewidth}{0.4pt} 

\noindent\textit{Feature definitions:}  
\begin{itemize}
   \item \textbf{Degree} ($k_i$): Represents the number of edges connected to node $i$, indicating its connectivity in the network.

   \item \textbf{Betweenness centrality} ($BC_i$): 
$BC_i = \sum_{s \neq i \neq t} \frac{\sigma_{st}(i)}{\sigma_{st}}$,  
where $\sigma_{st}$ is the total number of shortest paths between nodes $s$ and $t$, and $\sigma_{st}(i)$ is the number of those paths that pass through node $i$. 
Measures the fraction of shortest paths passing through node $i$, reflecting its potential control over information flow.
 
    \item \textbf{Normalised gene expression} ($z_i$): $z_i = \frac{x_i - \mu}{\sigma}$, where $x_i$ is expression value, $\mu$ mean, $\sigma$ standard deviation.  
    \item \textbf{Node role}: TF = transcription factor; Target = regulated gene; TF+Target = both.  
\end{itemize}
\label{tab:gnn_graph_structure}
\end{table}

A Graph Attention Network v2 (GATv2) architecture was employed.
It enhances neighbour aggregation by dynamically re-weighting edges using both node and edge features, and this enables finer capture of local topological variations \cite{brody2022gatv2}. Hyperparameters were tuned via Optuna, a framework for efficient automated search \cite{akiba2019optuna}.

\subsection{Model Evaluation}

\subsubsection{Primary Evaluation Metrics}

This study used two main evaluation metrics: the Area Under the Receiver Operating Characteristic Curve (AUROC) \cite{fawcett2006roc} and the Matthews Correlation Coefficient (MCC) \cite{matthews1975mcc}.

AUROC measures how well a model distinguishes between positive and negative cases by summarising the trade-off between \textit{sensitivity} (true positives correctly identified) and \textit{specificity} (true negatives correctly identified) across all classification thresholds \cite{fawcett2006roc}.MCC measures the general agreement between the predicted and actual labels, with a single value ranging from $-1$ (perfect disagreement) to $1$ (perfect agreement) \cite{matthews1975mcc}.

The AUROC and MCC were chosen as the main evaluation metrics for two reasons. First, AUROC is a widely accepted standard in biomedical classification, allowing direct comparison with prior studies \cite{fawcett2006roc}. Second, MCC provides a single summary score of agreement between predicted and true labels, giving equal weight to metastatic and primary classes in a perfectly balanced dataset \cite{matthews1975mcc}. Additional metrics (AUPRC \cite{saito2015auprc}, FDR (false discovery rate), FNR(false negative rate), FPR(false positive rate), MMCE(1- Accuracy), NPV (negative predictive value), PPV ( positive predictive value) are included in the appendix.

\subsubsection{Interpretation of Model Performance}

The performance of the model was interpreted by examining how well the samples with different metastatic statuses were separated in the data. For ML models, the principal component analysis (PCA) \cite{ringner2008pca} was applied to the top 100 to 1000 genes. PCA reduces high-dimensional gene expression data to a few components that capture most variance.  We examined whether ML models can separate patterns that PCA cannot. The PCA results also help explain the model performance, as accuracy may be related to the degree of separation seen in PCA.

For GNN models, combo plots of network-level features were used as a comparison. The features included clustering coefficients, degree variance, density, node counts, edge counts, triangles, and path length. This comparison shows the separability provided by simple graph statistics. We examined whether GNN models can identify additional topological patterns beyond these statistics. The results also help interpret model accuracy since stronger separation in these features may correspond to better performance.

\subsection{Sensitivity Analysis}
Sensitivity analysis evaluated the robustness of the top ML and GNN models by retraining them with optimal hyperparameters across seeds 2, 4, 6, 8, and 10.
For each seed, data were split 80/20 with stratification, the model was trained and AUROC was calculated on the test set. The results were visualised as bar graphs with mean and baseline lines to assess sensitivity to initialisation, supporting conclusions on the stability and reproducibility of the model.

\section{Results}

\begin{figure}[H]
  \centering
  \includegraphics[width=0.9\linewidth]{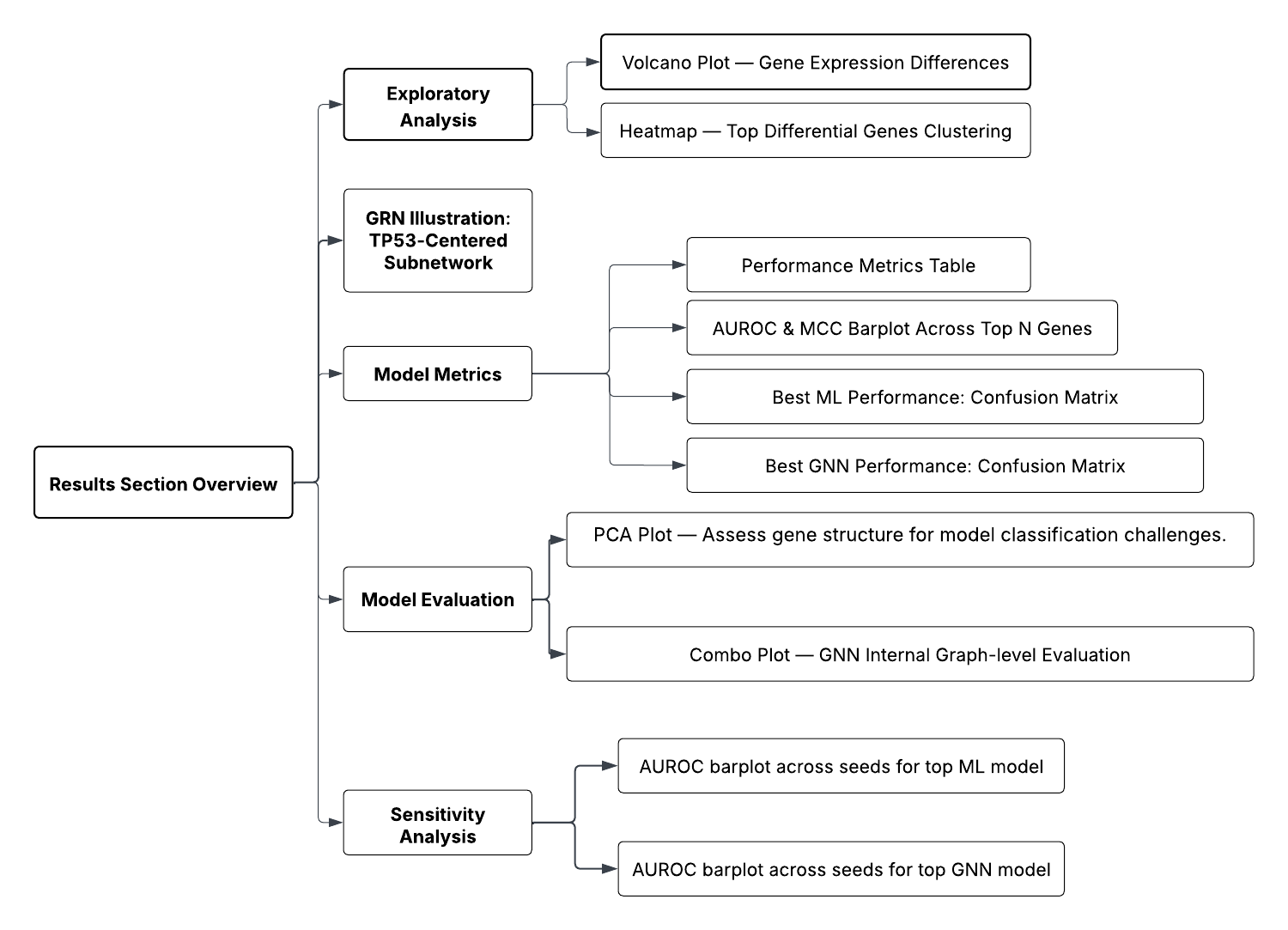}
  \caption{Results section overview.}
  \label{fig:results_overview}
\end{figure}

Figure~\ref{fig:results_overview} provides an overview of the Results section, structured into five components: 
(i) \textit{Exploratory Analysis} — volcano plot to assess whether systematic expression differences exist between metastatic and primary samples, and heatmap to explore if the top 50 genes show separation between the two phenotypes; 
(ii) \textit{GRN Illustration} — TP53-centered subnetwork example; 
(iii) \textit{Model Metrics} — performance table, AUROC/MCC barplots, and confusion matrices for the best ML and GNN models; 
(iv) \textit{Model Evaluation} — PCA to interpret overall model performance via sample separability by metastatic status, and combo plot to interpret graph-level patterns learned by the GNN; 
(v) \textit{Sensitivity Analysis} — AUROC barplots across seeds for the top ML and GNN models.

\subsection{Exploratory Analysis}
\subsubsection{Volcano Plot — Gene Expression Differences}
With the p-value threshold set at 0.05, the volcano plot showed that most genes reached significance. A subset of these genes was consistently upregulated or downregulated in metastatic samples, suggesting the presence of systematic expression differences, although the effect was modest.
\begin{figure}[H]
  \centering
  \includegraphics[width=0.9\linewidth]{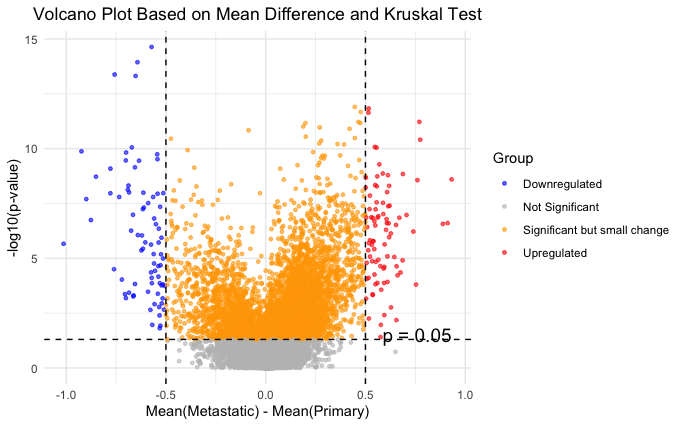}
  \caption{Volcano plot of genome-wide differential expression. Volcano plot comparing metastatic and primary samples. Most genes reached statistical significance (p $<$ 0.05), with subsets showing consistent upregulation or downregulation in metastatic samples, suggesting the presence of systematic expression differences.}
  \label{fig:volcano}
\end{figure}

\subsubsection{Heatmap — Top Differential Genes Clustering}

The heatmap of the top 50 differential genes (Kruskal, $|\text{mean diff}| > 0.5$) shows that metastatic and primary samples tend to group separately, although the separation is not complete. Some gene clusters have clear opposite expression patterns between the two groups, while others show mixed patterns, suggesting that only part of the genes display systematic differences related to metastatic status.

\begin{figure}[H]
  \centering
  \includegraphics[width=0.9\linewidth]{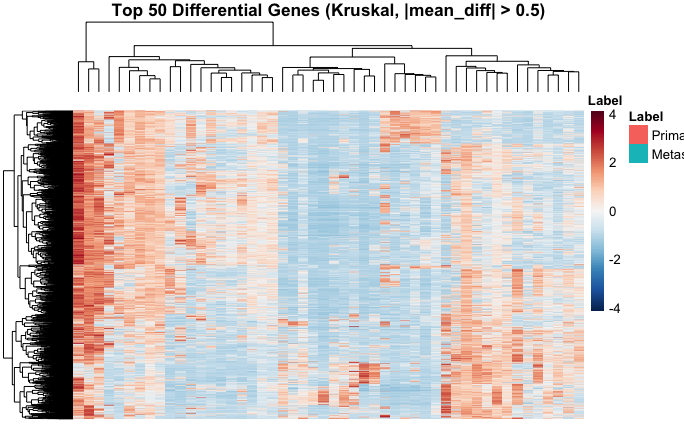}
  \caption{Heatmap of the top 50 differential genes (Kruskal–Wallis, $|\text{mean diff}| > 0.5$). Metastatic and primary samples tend to group separately, though not completely. Some clusters show opposite expression patterns, while others are mixed, indicating that only a subset of genes differs systematically with metastatic status.}
  \label{fig:heatmap}
  
\end{figure}

\begin{table}[H]
\centering
\label{tab:top10_genes}
\begingroup
\setstretch{1.0} 
\begin{tabular}{lcc}
\toprule
Gene & Mean Difference & p-value \\
\midrule
IRF1    & $-0.572$ & $2.31 \times 10^{-15}$ \\
CNN2    & $-0.643$ & $1.15 \times 10^{-14}$ \\
TGFB1   & $-0.757$ & $4.16 \times 10^{-14}$ \\
IL15RA  & $-0.652$ & $4.82 \times 10^{-14}$ \\
IGSF11  & $ 0.516$ & $1.48 \times 10^{-12}$ \\
\bottomrule
\end{tabular}
\caption*{\textbf{Top 5 genes ranked by p-value.} Among the top 5 genes, IRF1 \cite{irf1_ref}, CNN2 \cite{cnn2_ref}, and TGFB1 \cite{tgfb1_ref} are reported to be associated with metastasis. This supports the heatmap observation that certain clusters show systematic differences related to metastatic status.}
\endgroup
\end{table}

\subsection{Panda network illustration — TP53-Centered Subnetwork}

\begin{figure}[H]
  \centering
  \includegraphics[width=0.9\linewidth]{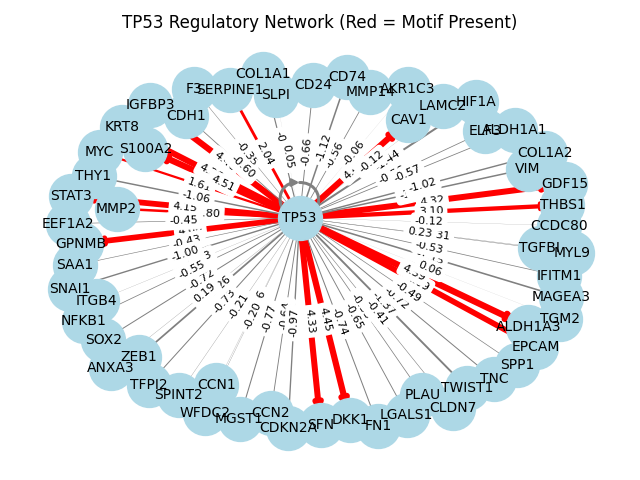}
  \caption{TP53-centered regulatory subnetwork.}
  \label{fig:tp53_network}
\end{figure}

As an illustrative example, Figure~\ref{fig:tp53_network} shows the TP53-centered regulatory subnetwork inferred from the PANDA network.  
Nodes represent genes and edges indicate regulatory connections, with red edges marking interactions from the TF–target prior. Edge labels show the corresponding PANDA regulatory strength values. 

\subsection{Model Metrics}
\subsubsection{Performance Metrics Table}

Table~\ref{tab:model_metrics_ml_8col} summarises the performance of traditional ML models (Random Forest, XGBoost, and ElasticNet) across four feature sets defined by the top 100, 200, 500, and 1000 most statistically significant differential genes. 
For all three algorithms, AUROC generally improved as the number of input genes increased, with the best results observed for XGBoost using the top 1000-gene set (AUROC = 0.7051, MCC = 0.2914). 
This trend suggests that traditional ML models benefit from a larger number of informative features within the evaluated range.

\begin{table}[H]
\centering
\caption{Performance metrics (AUROC and MCC) for traditional ML models across different numbers of top-ranked differential genes. 
Top N refers to the number of most significant differential genes selected based on statistical testing.}
\label{tab:model_metrics_ml_8col}
\resizebox{\textwidth}{!}{%
\begin{tabular}{lcccccccc}
\toprule
Model & \multicolumn{2}{c}{Top 100 Genes} & \multicolumn{2}{c}{Top 200 Genes} & \multicolumn{2}{c}{Top 500 Genes} & \multicolumn{2}{c}{Top 1000 Genes} \\
\cmidrule(lr){2-3} \cmidrule(lr){4-5} \cmidrule(lr){6-7} \cmidrule(lr){8-9}
& AUROC & MCC & AUROC & MCC & AUROC & MCC & AUROC & MCC \\
\midrule
RandomForest & 0.6601 & 0.2474 & 0.6665 & 0.2559 & 0.6872 & 0.2734 & 0.6911 & 0.2435 \\
XGBoost      & 0.6514 & 0.2629 & 0.6455 & 0.2244 & 0.6760 & 0.2148 & \textbf{0.7051} & \textbf{0.2914} \\
ElasticNet   & 0.6776 & 0.2450 & 0.6898 & 0.2545 & 0.6372 & 0.1278 & 0.6809 & 0.2431 \\
\bottomrule
\end{tabular}%
}
\end{table}

Table~\ref{tab:model_metrics_gnn} reports results for GNN models using the same feature sets. 
Unlike the traditional ML models, the GNN’s AUROC values show limited variation with feature set size. 
Although the top 500-gene set achieved the highest AUROC (0.6460), its MCC (0.1974) was notably lower than that of the top 100-gene set (0.2254). 
Given the marginal AUROC difference and substantially higher MCC, the top 100-gene configuration can be considered the best-performing GNN model in this comparison.


\begin{table}[H]
\setlength{\tabcolsep}{40pt}
\centering
\caption{Performance metrics (AUROC and MCC) for GNN models using different numbers of top-ranked differential genes. 
Top N indicates the number of most statistically significant differential genes included as input features.}
\label{tab:model_metrics_gnn}
\begin{tabular}{lcc}
\toprule
Model & AUROC & MCC \\
\midrule
GNN\_top100\_balanced  & \textbf{0.6423} & \textbf{0.2254} \\
GNN\_top200\_balanced  & 0.6196 & 0.1651 \\
GNN\_top500\_balanced  & 0.6460 & 0.1974 \\
GNN\_top1000\_balanced & 0.6367 & 0.2108 \\
\bottomrule
\end{tabular}
\end{table}

Overall, the AUROC of the top-performing GNN model (Top 100 genes)   remained lower than that of the best traditional ML model. 
This indicates that, for this dataset, increasing the number of differential genes benefits traditional ML models more substantially, whereas GNN performance is less sensitive to feature set size.

\subsubsection{AUROC \& MCC Barplot Across Top N Genes}

\begin{figure}[H]
\centering
\includegraphics[width=0.7\textwidth]{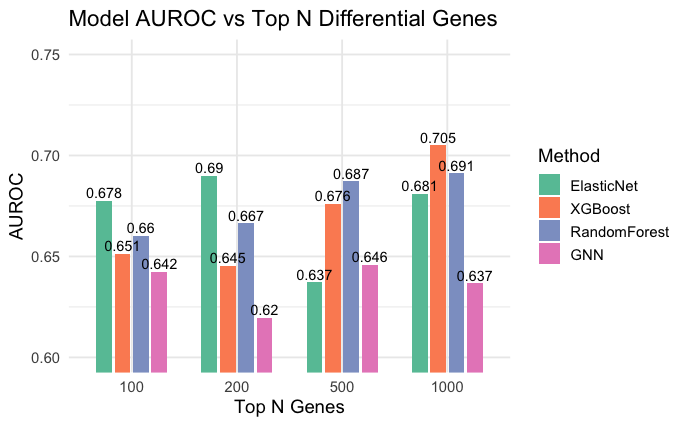}
\includegraphics[width=0.7\textwidth]{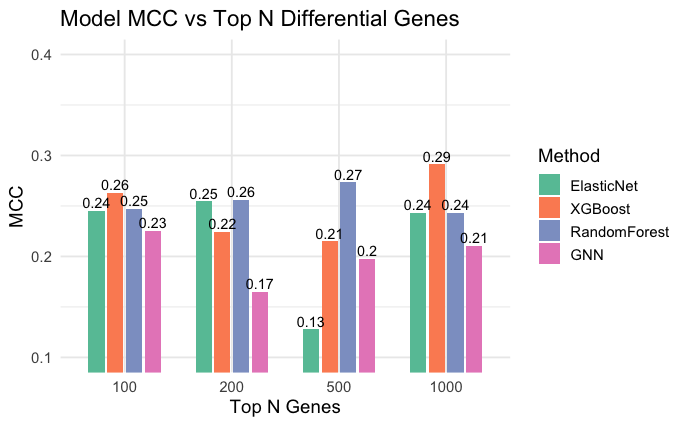}
\caption{AUROC and MCC trends across top-ranked gene sets for ML and GNN models}
\label{fig:ml_gnn_topN}
\end{figure}

Figure~\ref{fig:ml_gnn_topN} further illustrates these patterns by showing AUROC (top panel) and MCC (bottom panel) trends in the top N values of the four models. AUROC increases steadily for traditional ML methods, whereas GNN remains relatively flat. MCC results highlight that the best performance of GNN occurs at 100 genes, after which it declines. These findings indicate that the predictive performance of GNN may depend more on the quality of the features than the quantity, while traditional ML models gain from a larger set of features.

\subsubsection{Confusion Matrix for the Best-Performing ML Model}

\begin{figure}[H]
\centering
\includegraphics[width=0.8\textwidth]{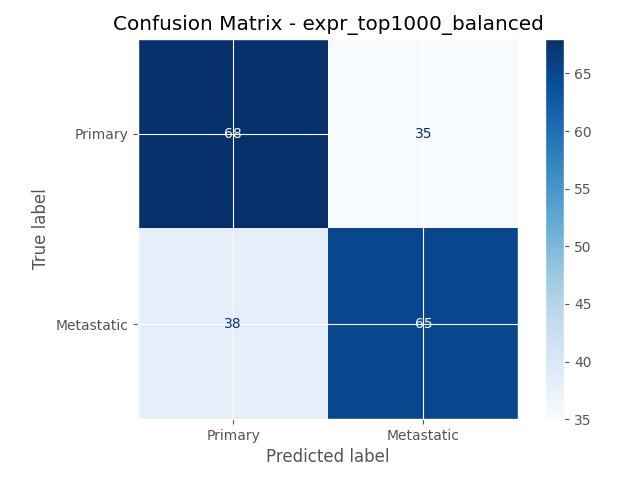}
\caption{Confusion matrix of XGBoost with top 1000 ranked differential genes.}
\label{fig:confmat_xgb_top1000}
\end{figure}

The traditional ML model with the highest performance was XGBoost, using the top 1,000 differential genes ranked. 
Figure~\ref{fig:confmat_xgb_top1000} shows its confusion matrix. The model achieved balanced performance across the two classes, correctly identifying
68 primary and 65 metastatic cases, while misclassifying 35 primary and 38 metastatic samples. This balance suggests that the model captures discriminative patterns in both classes, although there remains room for improvement in reducing false positives and false negatives.

\subsubsection{Confusion Matrix for the Best-Performing GNN Model}

\begin{figure}[H]
\centering
\includegraphics[width=0.8\textwidth]{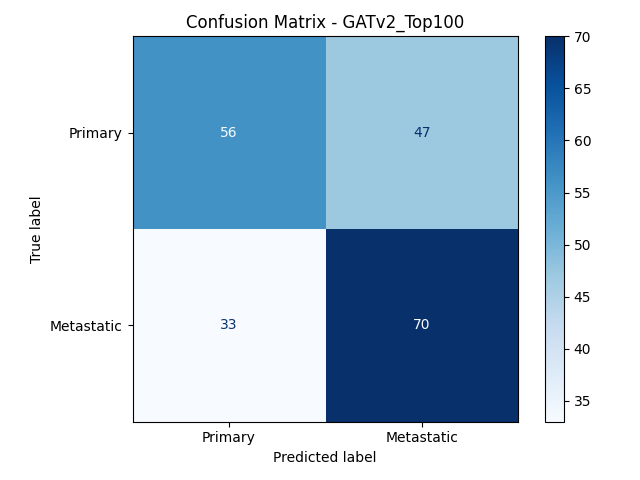}
\caption{Confusion matrix of GATv2 with top 100 ranked differential genes.}
\label{fig:confmat_gnn_top100}
\end{figure}

The best performing GNN configuration was GATv2 trained on the top 100 differential genes ranked. 
Figure~\ref{fig:confmat_gnn_top100} shows its confusion matrix. The model correctly classified 56 primary and 70 metastatic cases, while misclassifying 47 primary and 33 metastatic samples. Compared to the best ML model, the GNN shows slightly lower accuracy for the primary class but comparable or better performance for the metastatic class. 
Despite a lower AUROC than traditional ML, GNN may be more effective in identifying metastatic cases, which are often clinically critical.

\subsection{Model Evaluation}

\subsubsection{PCA Plot — Model Space Separability}

\begin{figure}[H]
\centering
\includegraphics[width=0.9\textwidth]{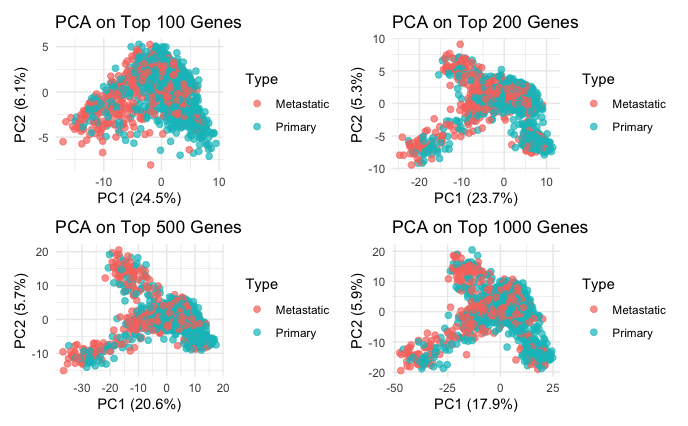}
\caption{PCA projections of top 100, 200, 500, and 1000 ranked differential genes, colored by metastasis status.}
\label{fig:pca_topN}
\end{figure}

Principal component analysis (PCA) was applied to the 100, 200, 500, and 1000 differential genes ranked first to evaluate the separability of the primary and metastatic samples (Figure~\ref{fig:pca_topN}).
Across all size sets, the two classes exhibit substantial overlap in the first two components, with no clear decision boundary.

This limited intrinsic separability helps explain the moderate AUROC values (0.60–0.70) for most models. The AUROC of the XGBoost model of 0.71 in the top 1,000 genes is therefore notable, indicating its ability to capture nonlinear patterns that are not evident in PCA.

Slightly better clustering is observed in the Top 200 and Top 500 gene sets, while the Top 100 set shows moderate overlap. For the Top 1000 genes, class overlap increases further, likely due to the inclusion of non-informative genes that dilute signal with noise. These results suggest that modest improvements in AUROC for traditional ML models with increasing top N are unlikely to be driven by strong intrinsic separability and instead may reflect the model’s capacity to capture more complex, nonlinear relationships beyond PCA. 

\subsubsection{Combo Plot — GNN Internal Graph-level Evaluation}

To investigate the reasons behind the modest AUROC of the GNN in the Top 100 genes, a combination graph was used to assess structural characteristics at the graph level. PCA in the same gene set indicated limited separability between primary and metastatic samples, suggesting a weak intrinsic signal in the expression space. 

Seven metrics were compared between sample-specific graphs: average clustering coefficient, degree variance, maximum clustering coefficient, number of edges, number of nodes, density, and number of triangles. 

Most metrics exhibited substantial overlap between the two classes, indicating a broadly similar global topology. Slight differences in degree variance, number of edges, and number of triangles suggest the presence of subtle local structural variations.

These findings indicate that the limited performance of the GNN is likely due to the absence of strong global structural separation, with only fine-grained local patterns available for learning.

\begin{figure}[H]
\centering
\includegraphics[width=0.85\textwidth]{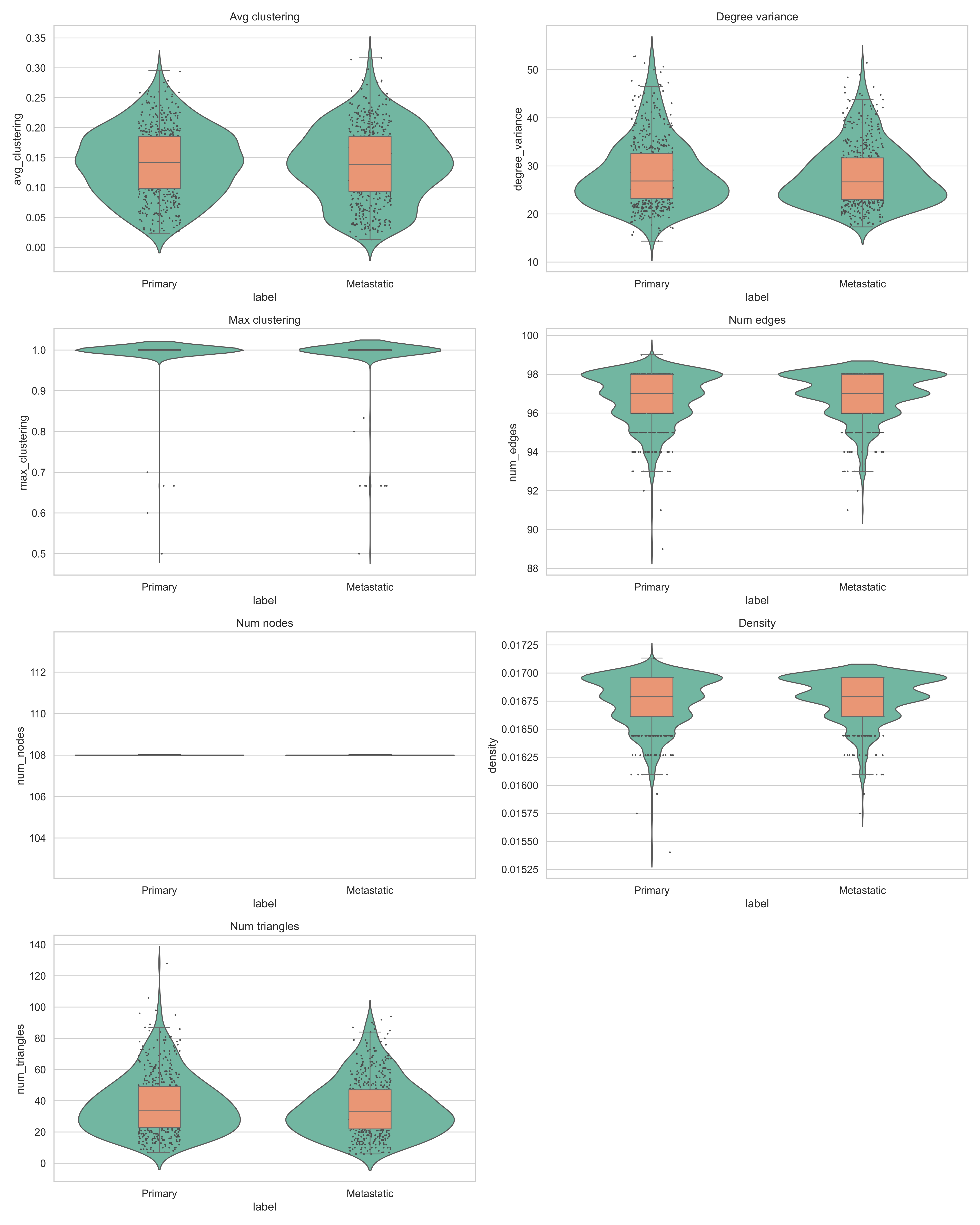}
\caption{Combo plot of graph-level features for the top 100 genes. Most metrics overlap between groups, with slight differences in degree variance, edges, and triangles, suggesting weak global separation and subtle local patterns, which may explain the GNN’s poor performance.}
\label{fig:combo_top100}
\end{figure}

\subsection{Sensitivity Analysis-Model Robustness to Random Seeds}

\begin{figure}[H]
    \centering
    \includegraphics[width=0.8\textwidth]{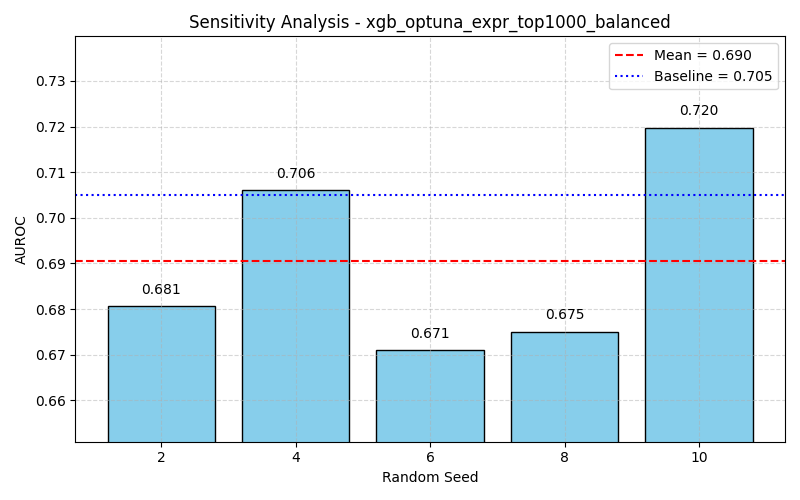}
    \caption{Sensitivity analysis of XGBoost (Top 1000 genes) with varying random seeds for data partitioning.}
    \label{fig:sensitivity_xgb_top1000}
\end{figure}

Sensitivity analysis for the XGBoost model trained on the Top 1000 genes was performed by varying the random seed for data partitioning across five runs. AUROC values ranged from 0.671 to 0.720, with a mean of 0.690. Moderate variation indicates that performance remains largely stable across different data partitions, with minor fluctuations suggesting a limited but manageable influence of sample composition on model outcomes.

\begin{figure}[H]
    \centering
    \includegraphics[width=0.8\textwidth]{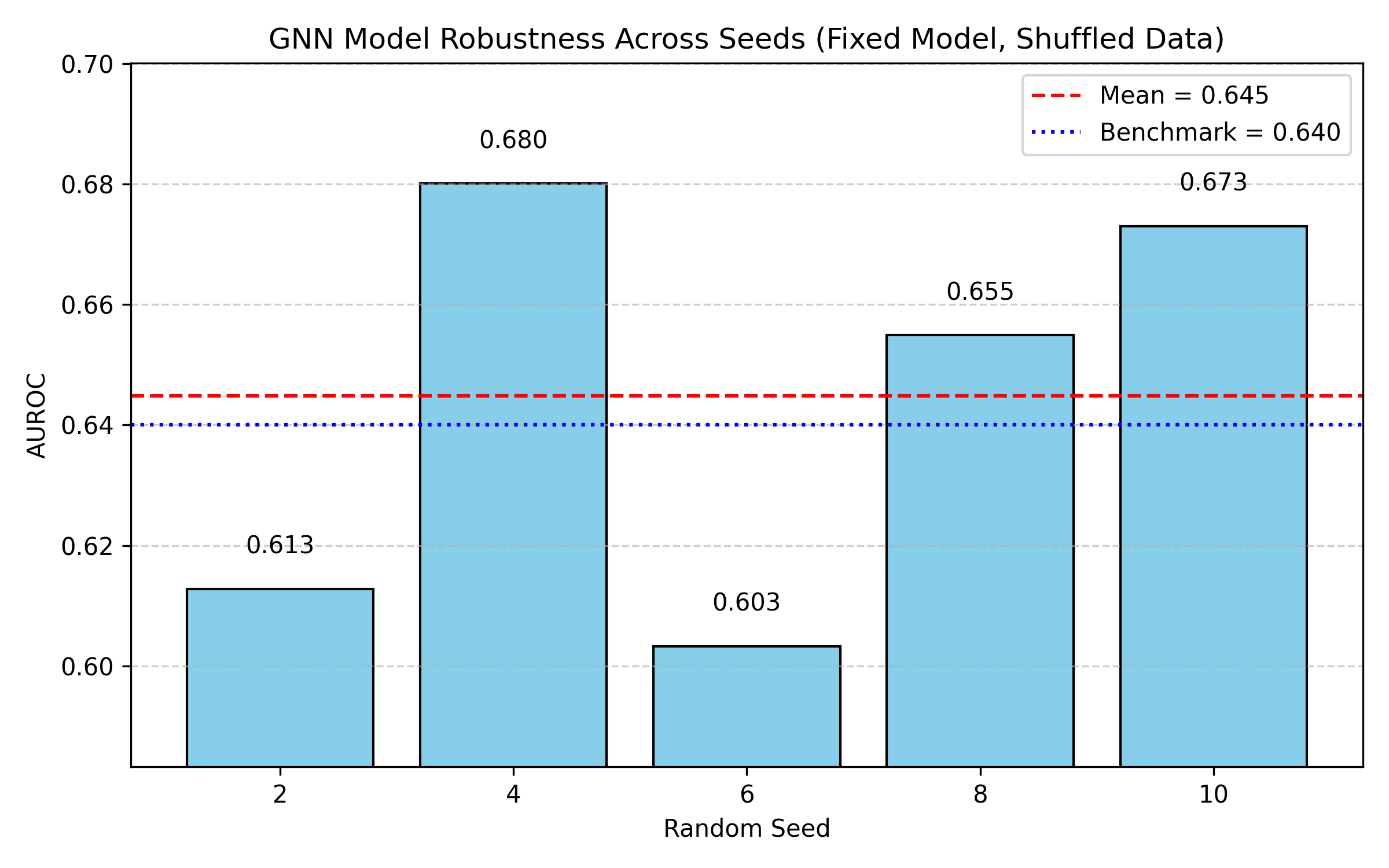}
    \caption{Robustness analysis of the GNN model (fixed architecture, shuffled data) across five random seeds.}
\end{figure}
Sensitivity analysis for the GNN model trained on the top 100 genes was also conducted. AUROC values ranged from 0.603 to 0.680, with a mean of 0.645, close to the benchmark of 0.640. The variation between seeds is moderate, indicating that the performance of the GNN is generally stable under different data partitions. Fluctuations suggest a limited but manageable influence on sample composition, emphasising the robustness of the model despite minor sensitivity to data splits.

\section{Discussion}

\paragraph{Summary}

This study explored two complementary strategies for predicting pancancer metastases. Common machine learning models were applied to CCLE gene expression data to evaluate predictive performance, assess feasibility in low- and middle-income countries (LMICs), and serve as benchmarks. In parallel, personalised gene regulatory networks were constructed using PANDA and LIONESS and incorporated into a graph neural network framework to capture patient-specific patterns for the prediction of metastases, with potential relevance for precision medicine.

The best traditional ML model (XGBoost with the top 1,000 genes) achieved an AUROC of 0.7051. The best GNN (top 100 genes) achieved an MCC of 0.2254 and an AUROC of 0.6423. Exploratory and PCA analyses showed a large overlap between classes in the expression space. The graph-level evaluation found similar global topologies with only small local differences. This suggests that traditional ML benefits from larger feature sets and captures non-linear patterns beyond intrinsic data separability. GNN performance is limited by weak network structural signals. Sensitivity analyses confirmed that both methods had stable results in random seeds, showing robustness to data partitioning.

\paragraph{Study Advantage}
Compared to previous work, this study demonstrates three key advantages. 

First, the XGBoost classifier achieved an AUROC of approximately 0.71 using a carefully filtered gene set, indicating its potential for low-cost preliminary detection of metastases in global health settings.

Second, our study builds sample-specific PANDA/LIONESS regulatory networks. Each patient has a unique network as input, and the network structure itself varies between individuals. In contrast, most existing GNN-based biomedical studies use a single fixed network topology with patient-specific node features \cite{emogi2021nmi, schulte2021glrp, gnnsubnet2022bioinformatics, fgcnsurv2023bioinformatics, mtgcn2022bib}. In these Graph-CNN studies, all patients share the same global network. Patient-specific subnetworks appear only in the interpretation stage, by selecting parts of this global network that influence predictions \cite{schulte2021glrp}.

Third, this study uniquely integrates both global ML benchmarking and personalised GNN modelling within the same framework, enabling analysis at both population and individual levels, an approach rarely seen in the literature.

\paragraph{Global health implication}
This study also has important implications for global health.

First, the ML component uses a limited set of selected differential genes, so full genome-wide sequencing is not required. This improves feasibility in resource-limited settings and allows deployment in decentralised laboratories \cite{teo2019global}. The framework is cancer-type agnostic, with a modular design that can be adapted to other cancers and metastatic sites \cite{li2022xgboost, jung2023breast}, ensuring scalability across populations and healthcare systems.

Secondly, the GNN used patient-specific regulatory networks from LIONESS-on-PANDA, integrating a TF–target prior with gene expression data \cite{lioness}. It allows for precision-level analysis in diverse populations, thereby fostering the advancement of personalised diagnosis and treatment.

Third, our dual framework combines population-level ML with patient-specific GNN. The ML approach demonstrates feasibility for use in training and policy contexts, while the GNN approach uses personalised gene regulatory networks to model patient-specific patterns in the prediction of metastases. Together, these complementary strategies could be integrated into cancer registries and screening systems to inform data-driven resource allocation \cite{hanna2020global, ascopubs_lmic}.

\paragraph{Limitation}
The limitations of this study are as follows. First, gene selection was based on the Kruskal–Wallis test with p-value ranking, a standard nonparametric method in gene expression analysis \cite{conesa2016survey}, but without effect size measures such as logarithmic fold change, which are widely recommended \cite{love2014moderated}. A more robust approach would combine statistical significance and effect size, for example, selecting genes with adjusted \(p\) value $<0.05$ and absolute $\log_{2}$ fold change above a predefined threshold. Second, the AUROC and MCC results for the GNN and traditional ML models were displayed together, which may imply a direct comparison. Although it is not strictly rigorous, this was intended only to benchmark the feasibility. Third, the GNN did not use subgraphs centred on nine key transcription factors: TP53, MYC, STAT3, HIF1A, NFKB1, SOX2, TWIST1, SNAI1, and ZEB1 \cite{yang2004,batlle2000,spaderna2008,yu2009,semenza2002,boumahdi2020,cooper_ras,muller2009,snail_nfkb} but instead trained on the entire network of the top 100 to 1000 ranked genes. This broader set may have introduced noise, masking topology signals. Fourth, the GNN used only expression-derived node features without additional biological labels, and this may have limited its ability to capture more complex topological patterns. Fifth, the models were trained and evaluated on a balanced dataset, and this can reduce robustness and generalisability because class imbalance is a well-known challenge in biomedical machine learning \cite{johnson2019survey}.

\paragraph{Future Work}

Future research should address the limitations of the study.

First, gene selection should combine statistical significance with effect size, using adjusted \(p\) values and \(\log_2\) fold change thresholds to improve biological relevance \cite{love2014moderated, conesa2016survey}. 
For example, genes with an adjusted \(p < 0.05\) and an absolute \(\log_2\) fold change greater than 1 could be filtered, ensuring that the selected genes are statistically significant and have biologically meaningful expression differences.

Second, constructing GNNs on biologically informed subgraphs, such as modules around TP53, STAT3, and other core transcription factors, can reduce noise and enhance the detection of topology signals. Subgraph-based methods such as GNN-SubNet \cite{gnnsubnet2022bioinformatics}, EMOGI \cite{emogi2021nmi}, and attention-based CGMega \cite{cgmmega2023} demonstrate the benefits of this approach.

Third, enriching node features with multiomics and clinical annotations could improve model performance. In practice, additional omics layers, such as DNA methylation, copy number variation, and mutation profiles, can be linked to the corresponding genes or patients. Clinical variables, such as age, tumour stage, and treatment history, can also be linked in the same way. These features can then be combined with expression values to create composite node feature vectors. Frameworks like FGCNSurv \cite{fgcnsurv2023bioinformatics}, MTGCN \cite{mtgcn2022bib}, and recent multimodal integration reviews \cite{multimodal2023review} support the value of data fusion.


\newpage
\bibliographystyle{plain}

\begin{thebibliography}{99}

\bibitem{who_cancer}
World Health Organization. Cancer. Available from: \url{https://www.who.int/news-room/fact-sheets/detail/cancer}. Accessed August 11, 2025.

\bibitem{ascopubs_lmic}
Sung H, Ferlay J, Siegel RL, Laversanne M, Soerjomataram I, Jemal A, Bray F. Global cancer statistics 2020: GLOBOCAN estimates of incidence and mortality worldwide for 36 cancers in 185 countries. CA Cancer J Clin. 2021;71(3):209–249. doi:10.3322/caac.21660.

\bibitem{frontiers_ai_cancer}
Esteva A, Robicquet A, Ramsundar B, Kuleshov V, DePristo M, Chou K, Cui C, Corrado G, Thrun S, Dean J. A guide to deep learning in healthcare. Nat Med. 2019;25:24–29. doi:10.1038/s41591-018-0316-z.

\bibitem{li2022xgboost}
Li X, Sun Z, He Y, et al.
XGBoost-based and tumor-immune characterized gene signature for the prediction of metastatic status in breast cancer.
J Transl Med. 2022;20:177.
doi:10.1186/s12967-022-03369-9.


\bibitem{jung2023breast}
Identification of breast cancer metastasis markers from gene expression profiles using machine learning approaches.
Genes (Basel). 2023;14(9):1820.
doi:10.3390/genes14091820.

\bibitem{kipf2017semi}
Kipf TN, Welling M. Semi-supervised classification with graph convolutional networks. In: 5th International Conference on Learning Representations (ICLR); 2017 Apr 24–26; Toulon, France. Available from: \url{https://openreview.net/forum?id=SJU4ayYgl}.

\bibitem{hamilton2017inductive}
Hamilton WL, Ying R, Leskovec J. Inductive representation learning on large graphs. In: Advances in Neural Information Processing Systems 30 (NeurIPS 2017); 2017 Dec 4–9; Long Beach, CA. p. 1024–1034. Available from: \url{https://proceedings.neurips.cc/paper/2017/hash/5dd9db5e033da9c6fb5ba83c7a7ebea9-Abstract.html}.

\bibitem{emogi2021nmi}
Schulte-Sasse R, Budach S, Hnisz D, Marsico A.
Integration of multiomics data with graph convolutional networks to identify new cancer genes and their associated molecular mechanisms.
Nat Mach Intell. 2021;3:513--526.
doi:10.1038/s42256-021-00325-y.

\bibitem{schulte2021glrp}
Neil D, Briody J, Lacoste A, Sim A, Creed P, Saffari A.
Interpretable Graph Convolutional Neural Networks for Inference on Noisy Knowledge Graphs.
In: Machine Learning for Health (ML4H) Workshop at NeurIPS; 2018.
arXiv:1812.00279.
doi:10.48550/arXiv.1812.00279.


\bibitem{gnnsubnet2022bioinformatics}
Pfeifer B, Saranti A, Holzinger A. GNN-SubNet: disease subnetwork detection with explainable graph neural networks. Bioinformatics. 2022;38(Suppl\_2):ii120–ii126. doi:10.1093/bioinformatics/btac478.

\bibitem{cgmmega2023}
Li H, Han Z, Sun Y, Wang F, Hu P, Gao Y, Bai X, Peng S, Ren C, Xu X, Liu Z, Chen H, Yang Y, Bo X.
CGMega: explainable graph neural network framework with attention mechanisms for cancer gene module dissection.
Nat Commun. 2024;15:5997.
doi:10.1038/s41467-024-49979-9.

\bibitem{mtgcn2022bib}
Peng W, Tang Q, Dai W, Chen T.
Improving cancer driver gene identification using multi-task learning on graph convolutional network.
Brief Bioinform. 2021;22(6):bbab432.
doi:10.1093/bib/bbab432.

\bibitem{breastgnn2025}
Agyekum EA, Kong W, Ren Y-Z, Issaka E, Baffoe J, Wang X, Tan G, Xiong C, Wang Z, Qian X, Shen X-J. A comparative analysis of three graph neural network models for predicting axillary lymph node metastasis in early-stage breast cancer. Sci Rep. 2025;15:13918. doi:10.1038/s41598-025-97257-z.

\bibitem{nsclcgnn2024}
Ju H, Kim K, Kim BI, Woo SK. Graph neural network model for prediction of non-small cell lung cancer lymph node metastasis using protein–protein interaction network and 18F-FDG PET/CT radiomics. Int J Mol Sci. 2024;25(2):698. doi:10.3390/ijms25020698.

\bibitem{crlmgnn2025}
Zhu Y, Yang W, Li Z, Pan C, Qi H. STG: Spatiotemporal Graph Neural Network with Fusion and Spatiotemporal Decoupling Learning for Prognostic Prediction of Colorectal Cancer Liver Metastasis. arXiv preprint arXiv:2505.03123. 2025 May 6.

\bibitem{panda}
Glass K, Huttenhower C, Quackenbush J, Yuan GC. Passing messages between biological networks to refine predicted interactions. PloS one. 2013 May 31;8(5):e64832.


\bibitem{lioness}
Kuijjer ML, Tung MG, Yuan G, Quackenbush J, Glass K. Estimating sample-specific regulatory networks. Iscience. 2019 Apr 26;14:226-40.

\bibitem{ccle}
Ghandi M, Huang FW, Jané-Valbuena J, Kryukov GV, Lo CC, McDonald III ER, Barretina J, Gelfand ET, Bielski CM, Li H, Hu K. Next-generation characterization of the cancer cell line encyclopedia. Nature. 2019 May 23;569(7757):503-8.


\bibitem{garcia2019dorothea}
Garcia-Alonso L, Holland CH, Ibrahim MM, Turei D, Saez-Rodriguez J. Benchmark and integration of resources for the estimation of human transcription factor activities. Genome research. 2019 Aug 1;29(8):1363-75.


\bibitem{yang2004}
Yang J, Mani SA, Donaher JL, Ramaswamy S, Itzykson RA, Come C, Savagner P, Gitelman I, Richardson A, Weinberg RA. Twist, a master regulator of morphogenesis, plays an essential role in tumor metastasis. cell. 2004 Jun 25;117(7):927-39.

\bibitem{batlle2000}
Batlle E, Sancho E, Francí C, Domínguez D, Monfar M, Baulida J, García de Herreros A. The transcription factor snail is a repressor of E-cadherin gene expression in epithelial tumour cells. Nature cell biology. 2000 Feb;2(2):84-9.

\bibitem{spaderna2008}
Spaderna S, Schmalhofer O, Wahlbuhl M, Dimmler A, Bauer K, Sultan A, Hlubek F, Jung A, Strand D, Eger A, Kirchner T. The transcriptional repressor ZEB1 promotes metastasis and loss of cell polarity in cancer. Cancer research. 2008 Jan 15;68(2):537-44.

\bibitem{yu2009}
Yu H, Kortylewski M, Pardoll D. Crosstalk between cancer and immune cells: role of STAT3 in the tumour microenvironment. Nature reviews immunology. 2007 Jan 1;7(1):41-51.


\bibitem{semenza2002}
Kenneth NS, Rocha S. Regulation of gene expression by hypoxia. Biochemical Journal. 2008 Aug 15;414(1):19-29.

\bibitem{boumahdi2020}
Boumahdi S, de Sauvage FJ. The great escape: tumour cell plasticity in resistance to targeted therapy. Nature reviews Drug discovery. 2020 Jan;19(1):39-56.

\bibitem{cooper_ras}
Sears RC. The life cycle of C-myc: from synthesis to degradation. Cell cycle. 2004 Sep 9;3(9):1131-5.

\bibitem{muller2009}
Muller PA, Vousden KH. Mutant p53 in cancer: new functions and therapeutic opportunities. Cancer cell. 2014 Mar 17;25(3):304-17.

\bibitem{snail_nfkb}
Wu Y, Deng J, Rychahou PG, Qiu S, Evers BM, Zhou BP. Stabilization of snail by NF-\(\kappa\)B is required for inflammation-induced cell migration and invasion. Cancer cell. 2009 May 5;15(5):416-28.


\bibitem{kruskal1952}
Kruskal WH, Wallis WA. Use of ranks in one-criterion variance analysis. Journal of the American statistical Association. 1952 Dec 1;47(260):583-621.


\bibitem{chawla2002smote}
Chawla NV, Bowyer KW, Hall LO, Kegelmeyer WP. SMOTE: synthetic minority over-sampling technique. Journal of artificial intelligence research. 2002 Jun 1;16:321-57.


\bibitem{he2009learning}
He H, Garcia EA. Learning from imbalanced data. IEEE Transactions on knowledge and data engineering. 2009 Jun 26;21(9):1263-84.


\bibitem{piccolo2022classify}
Piccolo SR, Mecham A, Golightly NP, Johnson JL, Miller DB. The ability to classify patients based on gene-expression data varies by algorithm and performance metric. PLoS computational biology. 2022 Mar 11;18(3):e1009926.



\bibitem{zou2005regularization}
Zou H, Hastie T. Regularization and variable selection via the elastic net. Journal of the Royal Statistical Society Series B: Statistical Methodology. 2005 Apr;67(2):301-20.


\bibitem{breiman2001random}
Breiman L. Random forests. Machine learning. 2001 Oct;45(1):5-32.


\bibitem{chen2016xgboost}
Chen T. XGBoost: A Scalable Tree Boosting System. Cornell University. 2016.

\bibitem{freeman1977}
Freeman LC. A set of measures of centrality based on betweenness. Sociometry. 1977 Mar 1:35-41.

\bibitem{newman2010}
Yang, S. \textit{Networks: An Introduction} by M.~E.~J.~Newman. Oxford, UK: Oxford University Press, 720~pp., \$85.00.


\bibitem{bishop2006prml}
Bishop CM, Nasrabadi NM. Pattern recognition and machine learning. New York: springer; 2006 Aug 17.


\bibitem{brody2022gatv2}
Brody S, Alon U, Yahav E. How attentive are graph attention networks?. arXiv preprint arXiv:2105.14491. 2021 May 30.

\bibitem{akiba2019optuna}
Akiba T, Sano S, Yanase T, Ohta T, Koyama M. Optuna: A next-generation hyperparameter optimization framework. InProceedings of the 25th ACM SIGKDD international conference on knowledge discovery \& data mining 2019 Jul 25 (pp. 2623-2631).

\bibitem{fawcett2006roc}
Fawcett T. An introduction to ROC analysis. Pattern recognition letters. 2006 Jun 1;27(8):861-74.

\bibitem{matthews1975mcc}
Matthews BW. Comparison of the predicted and observed secondary structure of T4 phage lysozyme. Biochimica et Biophysica Acta (BBA)-Protein Structure. 1975 Oct 20;405(2):442-51.


\bibitem{saito2015auprc}
Saito T, Rehmsmeier M. The precision-recall plot is more informative than the ROC plot when evaluating binary classifiers on imbalanced datasets. PloS one. 2015 Mar 4;10(3):e0118432.



\bibitem{ringner2008pca}
Ringnér M. What is principal component analysis?. Nature biotechnology. 2008 Mar;26(3):303-4.



\bibitem{irf1_ref} 
Hong M, Zhang Z, Chen Q, Lu Y, Zhang J, Lin C, Zhang F, Zhang W, Li X, Zhang W, Li X. IRF1 inhibits the proliferation and metastasis of colorectal cancer by suppressing the RAS-RAC1 pathway. Cancer management and research. 2018 Dec 31:369-78.

\bibitem{cnn2_ref} 
Bin X, Luo Y, Sun Z, Lin C, Huang P, Tu Z, Li L, Qu C, Long J, Zhou S. The role of H2-calponin antigen in cancer metastasis: presence of autoantibodies in liver cancer patients. International Journal of Molecular Sciences. 2023 Jun 7;24(12):9864.

\bibitem{tgfb1_ref} 
Katsuno Y, Lamouille S, Derynck R. TGF-$\beta$ signaling and epithelial–mesenchymal transition in cancer progression. Current opinion in oncology. 2013 Jan 1;25(1):76-84.

\bibitem{teo2019global}
Tekola-Ayele F, Rotimi CN. Translational genomics in low-and middle-income countries: opportunities and challenges. Public health genomics. 2015 Jun 26;18(4):242-7.

\bibitem{hanna2020global}
Allemani C, Matsuda T, Di Carlo V, Harewood R, Matz M, Nikšić M, Bonaventure A, Valkov M, Johnson CJ, Estève J, Ogunbiyi OJ. Global surveillance of trends in cancer survival 2000–14 (CONCORD-3): analysis of individual records for 37 513 025 patients diagnosed with one of 18 cancers from 322 population-based registries in 71 countries. The Lancet. 2018 Mar 17;391(10125):1023-75.


\bibitem{conesa2016survey}
Conesa A, Madrigal P, Tarazona S, Gomez-Cabrero D, Cervera A, McPherson A, Szcześniak MW, Gaffney DJ, Elo LL, Zhang X, Mortazavi A. A survey of best practices for RNA-seq data analysis. Genome biology. 2016 Jan 26;17(1):13.



\bibitem{love2014moderated}
Love MI, Huber W, Anders S. Moderated estimation of fold change and dispersion for RNA-seq data with DESeq2. Genome biology. 2014 Dec 5;15(12):550.


\bibitem{johnson2019survey}
Johnson JM, Khoshgoftaar TM. Survey on deep learning with class imbalance. Journal of big data. 2019 Dec;6(1):1-54.


\bibitem{fgcnsurv2023bioinformatics}
Wen G, Li L. FGCNSurv: dually fused graph convolutional network for multi-omics survival prediction. Bioinformatics. 2023 Aug 1;39(8):btad472.

\bibitem{multimodal2023review}
Wen H, Ding J, Jin W, Wang Y, Xie Y, Tang J. Graph neural networks for multimodal single-cell data integration. InProceedings of the 28th ACM SIGKDD conference on knowledge discovery and data mining 2022 Aug 14 (pp. 4153-4163).






\end{thebibliography}

\end{document}